\PassOptionsToClass{nonacm}{acmart}
\documentclass[sigconf]{acmart}

\AtBeginDocument{%
  }

\usepackage{algorithm}
\usepackage{algorithmic}
\usepackage{booktabs}
\usepackage{multirow}
\usepackage[table]{xcolor}
\usepackage{pifont}
\definecolor{bestgreen}{HTML}{C1E6C0}
\definecolor{failred}{HTML}{F4CCCC}

\begin{document}

\title{POET: Power-Oriented Evolutionary Tuning for LLM-Based RTL PPA Optimization}

\author{Heng Ping}
\email{hping@usc.edu}
\affiliation{%
  \institution{University of Southern California}
  \country{United States}}

\author{Peiyu Zhang}
\email{pzhang65@usc.edu}
\affiliation{%
  \institution{University of Southern California}
  \country{United States}}

\author{Zhenkun Wang}
\email{zhenkunw@usc.edu}
\affiliation{%
  \institution{University of Southern California}
  \country{United States}}

\author{Shixuan Li}
\email{sli97750@usc.edu}
\affiliation{%
  \institution{University of Southern California}
  \country{United States}}

\author{Anzhe Cheng}
\email{anzheche@usc.edu}
\affiliation{%
  \institution{University of Southern California}
  \country{United States}}

\author{Wei Yang}
\email{wyang930@usc.edu}
\affiliation{%
  \institution{University of Southern California}
  \country{United States}}

\author{Paul Bogdan}
\authornote{Corresponding author.}
\email{pbogdan@usc.edu}
\affiliation{%
  \institution{University of Southern California}
  \country{United States}}

\author{Shahin Nazarian}
\authornotemark[1]
\email{shahin.nazarian@usc.edu}
\affiliation{%
  \institution{University of Southern California}
  \country{United States}}

\renewcommand{\shortauthors}{Ping et al.}

\begin{abstract}
Applying large language models (LLMs) to RTL code optimization for improved power, performance, and area (PPA) faces two key challenges: ensuring functional correctness of optimized designs despite LLM hallucination, and systematically prioritizing power reduction within the multi-objective PPA trade-off space. We propose POET (Power-Oriented Evolutionary Tuning), a framework that addresses both challenges. For functional correctness, POET introduces a differential-testing-based testbench generation pipeline that treats the original design as a functional oracle, using deterministic simulation to produce golden references and eliminating LLM hallucination from the verification process. For PPA optimization, POET employs an LLM-driven evolutionary mechanism with non-dominated sorting, power-first intra-level ranking, and proportional survivor selection to steer the search toward the low-power region of the Pareto front without manual weight tuning. Evaluated on the RTL-OPT benchmark across 40 diverse RTL designs, POET achieves 100\% functional correctness, the best power on all 40 designs, and competitive area and delay improvements.
\end{abstract}




\maketitle

The rapid advancement of large language models (LLMs) \cite{yang2026auditing,yang2025toward,li2025climatellm} is reshaping electronic design automation (EDA), enabling new capabilities across design space exploration, verification, and optimization~\cite{zang2025dawn,xu2025llmeda,pan2025survey}. In particular, the strong code generation abilities of LLMs have inspired growing interest in RTL design~\cite{yang2025llmverilog}, and benchmarks such as VerilogEval~\cite{liu2023verilogeval,pinckney2025revisiting} and RTLLM~\cite{lu2024rtllm, liu2024openllm} have further accelerated this area. However, existing efforts~\cite{ping2025hdlcore,ping2025verimoa} predominantly target \textit{spec-to-RTL generation}, i.e., producing functionally correct Verilog from specification, while paying limited attention to PPA quality. In practice, initial RTL implementations are seldom PPA-optimal regardless of whether they are written by humans or generated by LLMs~\cite{hsin2026evolve}. \textit{RTL-to-RTL optimization}, which transforms a functionally correct design into a PPA-superior equivalent, therefore constitutes an important yet underexplored direction~\cite{delorenzo2024make}. Meanwhile, the proliferation of IoT devices, mobile computing, and embedded systems has made low-power optimization an increasingly essential objective in modern chip design~\cite{chandrakasan2008next}.

Applying LLMs to power-centric RTL PPA optimization faces two fundamental challenges. The first concerns the \textit{functional correctness of optimized designs}. Due to inherent hallucination tendencies, LLMs frequently produce functional errors when generating PPA-optimized variants~\cite{abdelatty2025pluto}, with experiments on RTL-OPT~\cite{lu2026benchmark} showing that even DeepSeek-R1 introduces errors in over 25\% of attempts. Since functional incorrectness renders any PPA improvement meaningless, robust verification is essential. Existing approaches employ diverse verification strategies: VeriOpt~\cite{tasnia2025veriopt} and LLM-VeriPPA~\cite{thorat2025llmverippa} rely on manually crafted testbenches, which are labor-intensive and difficult to scale; SymRTLO~\cite{wang2025symrtlo} directs the LLM to generate testbenches from the input RTL, yet the results often suffer from low reliability; although ASPEN~\cite{zhang2025aspen} employs equivalence checking, its simplified formulation with sparse constraints cannot fully guarantee functional equivalence. These limitations call for a more reliable and automated verification method.

The second challenge lies in the \textit{directionality and effectiveness of power-prioritized PPA optimization}. PPA metrics exhibit fundamental trade-offs, forming a Pareto front where improving one objective necessarily degrades another~\cite{gubbi2025prompting}. Existing methods handle this multi-objective nature inadequately. In-context learning approaches such as VeriOpt~\cite{tasnia2025veriopt} and LLM-VeriPPA~\cite{thorat2025llmverippa} embed optimization techniques into prompts but lack systematic design space exploration. Evolutionary approaches like REvolution~\cite{min2025revolution} and VFlow~\cite{wei2025vflow} collapse the multi-objective problem into a single weighted fitness score (e.g., $F = \alpha \cdot \Delta P + \beta \cdot \Delta A + \gamma \cdot \Delta T$), which requires manual weight tuning and cannot reach non-convex Pareto regions. None of these approaches can systematically prioritize power while maintaining Pareto optimality.

To address both challenges, we propose \textbf{POET} (Power-Oriented Evolutionary Tuning), a framework that integrates reliable functional verification with power-oriented multi-objective RTL PPA optimization. For functional correctness, POET employs a \textit{differential-testing-based testbench generation pipeline}. The key insight is that the input design, while PPA-suboptimal, is functionally correct and can serve as a functional oracle~\cite{lu2026benchmark}. POET directs the LLM to extract a functional specification and generate diverse test stimuli, which are then simulated through the original design to produce golden output signals. The resulting input-output pairs are assembled into a checking testbench, ensuring expected outputs come from deterministic simulation rather than LLM reasoning. For PPA optimization, POET adopts an evolutionary algorithm~\cite{qi2024evolution} that iteratively generates, evaluates, and selects RTL design variants through LLM-driven mutation and crossover. Designs are ranked via power-oriented non-dominated sorting, which first partitions the population into Pareto levels, then sorts each level by power in ascending order so that lower-power designs are always preferred among Pareto-equivalent candidates. 
A proportional survivor selection strategy then allocates more slots to higher-priority levels while preserving diversity from lower levels, steering the search toward the low-power region of the Pareto front without manually tuned weight parameters.

\noindent\textbf{Our key contributions:}
\begin{itemize}
    \item \textbf{Differential-Testing-Based Testbench Generation}: We propose an automated pipeline that treats the functionally correct input design as an oracle. Golden input-output pairs are obtained by simulating LLM-generated test stimuli through the original design, from which reliable checking testbenches are assembled, avoiding manual testbench creation or LLM-generated testbenches.

    \item \textbf{Power-Oriented Multi-Objective Evolutionary Optimization}: We introduce an LLM-driven evolutionary mechanism employing non-dominated sorting, power-first intra-level ranking, and proportional survivor selection to systematically prioritize power reduction while preserving Pareto optimality, without manual weight tuning.

    \item \textbf{Superior Performance}: Evaluation on the RTL-OPT benchmark across 40 diverse RTL designs demonstrates that POET achieves state-of-the-art power reduction with competitive area and delay improvements. Ablation studies further confirm the effectiveness of each component.
\end{itemize}

\section{Background}\label{sec:background}
\subsection{LLM-Based RTL Optimization}\label{subsec:llm-rtl-opt}

With the rapid progress of LLMs in complex tasks~\cite{chen2026self,yang2025maestro,yang2025learning,chang2025survey,ye2024domain, zhang2025mihc, duan2024structure}, a key prerequisite for RTL optimization is high-quality benchmarks. Unlike RTL code generation, where benchmarks are well established, dedicated datasets for RTL PPA optimization have only recently emerged. PfP~\cite{gubbi2025prompting} constructs a low-power RTL benchmark by pairing baseline modules with manually crafted low-power variants, though limited to small modules. RTL-OPT~\cite{lu2026benchmark} provides a larger-scale benchmark of 40 expert-crafted designs covering diverse categories. However, the limited scale of optimization data constrains training-based approaches such as PPA-RTL~\cite{zhao2025hardware}.

Given these limitations, most recent works adopt inference-time agentic frameworks combining LLMs with external tool feedback. In-context learning approaches such as VeriOpt~\cite{tasnia2025veriopt} and LLM-VeriPPA~\cite{thorat2025llmverippa} incorporate PPA feedback into iterative prompting but lack systematic design space exploration. Structured rewriting approaches take a different path: RTLRewriter~\cite{yao2024rtlrewriter} partitions RTL designs and employs cost-aware MCTS for rewriting; ASPEN~\cite{zhang2025aspen} combines LLM-guided rewrite rules with e-graph-based equality saturation; SymRTLO~\cite{wang2025symrtlo} integrates AST templates for datapath optimization with symbolic scripts for FSM minimization. These methods focus on localized transformations without a global multi-objective search strategy. A third line of work, including REvolution~\cite{min2025revolution}, EvolVE~\cite{hsin2026evolve}, and VFlow~\cite{wei2025vflow}, applies evolutionary algorithms to search more broadly by evolving populations of RTL candidates. However, as detailed in Section~\ref{subsec:evo-hw}, these approaches uniformly rely on weighted-sum fitness or proxy metrics, which cannot express power-centric optimization preferences without manual hyperparameter tuning.

\subsection{Evolutionary Approaches for RTL Design}\label{subsec:evo-hw}

Recent work has demonstrated the effectiveness of combining evolutionary computation (EC) with LLMs for RTL design~\cite{min2025revolution,hsin2026evolve,wei2025vflow}. In this paradigm, a population of design candidates is maintained and iteratively refined through an evolutionary loop of offspring generation, evaluation, and survivor selection. Each individual represents a design candidate containing the RTL implementation and its evaluation results. New offspring are produced by prompting the LLM with selected parents and an evolutionary operator such as mutation (improving a single parent) or crossover (combining two parents). Each offspring is evaluated for functional correctness and PPA quality, and the fittest ones survive into the next generation.

A critical design choice is how to define fitness for PPA optimization. REvolution~\cite{min2025revolution} computes a weighted sum $F = \alpha \cdot \Delta P + \beta \cdot \Delta A + \gamma \cdot \Delta T$ with manually set weights. EvolVE~\cite{hsin2026evolve} uses the area-delay product as a proxy metric. VFlow~\cite{wei2025vflow} decomposes the search into specialized populations (functionality-first, area-first, timing-first, and balanced), but each still optimizes a single scalar objective. Despite their broader design space exploration, these approaches all rely on weighted-sum scalarization or proxy metrics, preventing them from expressing a systematic preference for power reduction. In contrast, POET adopts multi-objective non-dominated sorting to preserve true Pareto optimality while injecting power preference through the selection mechanism, as detailed in Section~\ref{sec:method}.

\begin{figure*}[t]
\centering
\includegraphics[width=0.95\textwidth]{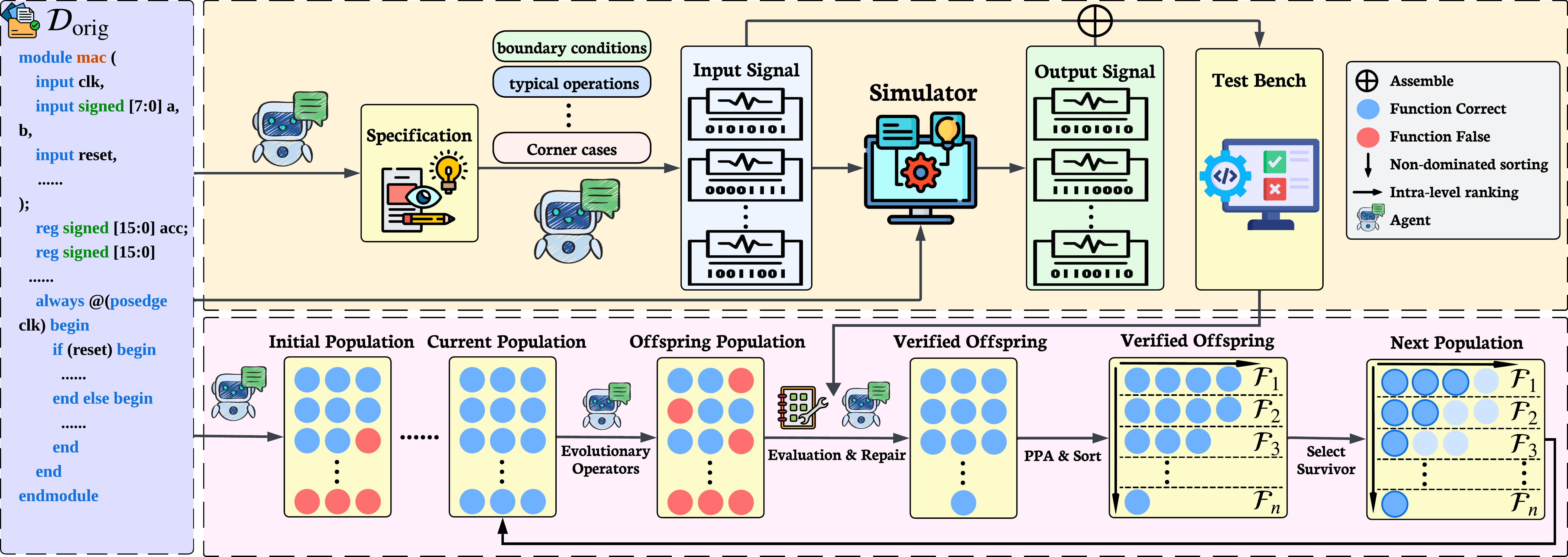}
\caption{POET. Top: differential-testing-based testbench generation. Bottom: power-oriented evolutionary optimization.}
\label{fig:overview}
\end{figure*}

\section{Method}\label{sec:method}

Figure~\ref{fig:overview} illustrates the overall architecture of POET, which comprises two core components: a differential-testing-based testbench generation pipeline that treats the original functionally correct design as a golden oracle to produce a reliable testbench, and a power-oriented evolutionary optimization engine that iteratively generates, evaluates, and selects RTL design variants.

\subsection{Problem Formulation}\label{subsec:formulation}

Given a functionally correct RTL design $\mathcal{D}_\text{orig}$ with PPA metrics $\mathbf{m}_\text{orig} = (P_\text{orig}, A_\text{orig}, D_\text{orig})$ for power, area, and critical path delay, the goal is to find a functionally equivalent design $\mathcal{D}^*$ with superior PPA quality:
\begin{equation}\label{eq:obj}
    \mathcal{D}^* = \arg\min_{\mathcal{D}} \; \mathbf{m}(\mathcal{D}) = (P, A, D)
    \quad \text{s.t.} \quad \mathcal{D} \equiv_f \mathcal{D}_\text{orig}
\end{equation}
Since PPA objectives are generally conflicting~\cite{gubbi2025prompting}, we adopt Pareto dominance to compare designs:
\begin{equation}\label{eq:dominance}
    \mathcal{D}_a \succ \mathcal{D}_b \iff \mathbf{m}(\mathcal{D}_a) \leq \mathbf{m}(\mathcal{D}_b) \;\land\; \mathbf{m}(\mathcal{D}_a) \neq \mathbf{m}(\mathcal{D}_b)
\end{equation}
A design is Pareto optimal if no other feasible design dominates it. To systematically prefer power reduction without collapsing into a single scalar, POET ranks designs on the same Pareto level (i.e., neither dominates the other) by power in ascending order:
\begin{equation}\label{eq:power_rank}
    \mathcal{D}_a \succ_P \mathcal{D}_b \iff P_a < P_b
\end{equation}
This steers the search toward the low-power region of the Pareto front while preserving multi-objective trade-off information, unlike weighted-sum formulations~\cite{min2025revolution,hsin2026evolve} that require manual tuning.

\subsection{Differential-Testing-Based Testbench Generation}\label{subsec:testbench}

Enforcing the functional equivalence constraint $\mathcal{D} \equiv_f \mathcal{D}_{\text{orig}}$ (Eq.~\ref{eq:obj}) throughout the evolutionary process requires a reliable testbench for every input design. Existing approaches either rely on simplified equivalence checking~\cite{zhang2025aspen}, which cannot fully guarantee functional equivalence, or assign testbench generation entirely to an LLM~\cite{wang2025symrtlo}, where hallucination-induced errors undermine verification reliability. POET addresses this through a differential testing pipeline that leverages LLMs for tasks where they excel, such as specification comprehension and test scenario design, while delegating output computation to deterministic simulation.

The key insight is to decompose testbench generation into input-side tasks, well-suited for LLMs, and output-side tasks, which require cycle-accurate reasoning that LLMs handle poorly for sequential circuits. By simulating $\mathcal{D}_{\text{orig}}$ with LLM-generated input stimuli, we obtain golden output signals without relying on LLM reasoning. The pipeline comprises four steps:

\noindent\textbf{Step~1: Specification Extraction.} The LLM analyzes $\mathcal{D}_{\text{orig}}$ and produces a structured specification $\mathcal{S}$ following a predefined template, including module interface, sequential/combinational classification, reset behavior, and functional description:
\begin{equation}\label{eq:spec}
    \mathcal{S} = \mathrm{LLM}_{\text{spec}}(\mathcal{D}_{\text{orig}})
\end{equation}
Extracting $\mathcal{S}$ as an intermediate representation, rather than generating test vectors directly from raw RTL, provides a concise functional summary that facilitates more accurate test scenario identification.

\noindent\textbf{Step~2: Test Vector Generation.} Based on $\mathcal{S}$, the LLM identifies test scenarios covering boundary conditions, typical operations, and corner cases, and generates the corresponding input signal sequences:
\begin{equation}\label{eq:vec}
    \mathbf{V} = \{v_1, v_2, \ldots, v_n\} = \mathrm{LLM}_{\text{vec}}(\mathcal{S})
\end{equation}
where each $v_i$ is a time-indexed input vector for one test scenario.

\noindent\textbf{Step~3: Golden Output Capture.} Rather than asking the LLM to reason about expected outputs, we construct a stimulus testbench from a predefined template that applies $\mathbf{V}$ to $\mathcal{D}_{\text{orig}}$ and records all output signals via an RTL simulator:
\begin{equation}\label{eq:golden}
    \mathbf{O} = \{o_1, o_2, \ldots, o_n\} = \mathrm{Sim}(\mathcal{D}_{\text{orig}},\, \mathbf{V})
\end{equation}
Since $\mathcal{D}_{\text{orig}}$ is functionally correct, $\mathbf{O}$ constitutes a reliable golden reference. This avoids potential hallucination that would arise from relying on LLM reasoning for output computation.

\noindent\textbf{Step~4: Checking Testbench Assembly and Validation.} The input-output pairs $(\mathbf{V}, \mathbf{O})$ are assembled into a checking testbench $\mathcal{T}$ through a predefined template:
\begin{equation}\label{eq:tb}
    \mathcal{T} = \mathrm{Assemble}(\mathbf{V},\, \mathbf{O})
\end{equation}
A validation step verifies that $\mathcal{D}_{\text{orig}}$ passes all assertions in $\mathcal{T}$; the testbench is accepted only upon successful validation. The validated $\mathcal{T}$ then serves as the functional verification oracle for all candidate designs, enforcing:
\begin{equation}\label{eq:verify}
    \mathcal{D} \equiv_f \mathcal{D}_{\text{orig}} \;\Leftarrow\; \mathcal{T}(\mathcal{D}) = \texttt{PASS}
\end{equation}
where $\mathcal{T}(\mathcal{D})$ denotes executing testbench on candidate design.

\subsection{Power-Oriented Evolutionary Optimization}\label{subsec:evolution}

RTL-to-RTL optimization requires selecting and combining design-specific rewriting strategies across a large transformation space that demands broad exploration, while synthesis feedback from each attempt should be exploited to guide subsequent efforts. Evolutionary algorithms are naturally suited for this task, as they maintain diverse candidates for exploration while using fitness-based selection for exploitation~\cite{min2025revolution}.

POET formulates the evolutionary process over a population of RTL design candidates. Each individual is defined as $\mathcal{I} = (\mathcal{D}, \mathbf{m})$, where $\mathcal{D}$ is the RTL implementation and $\mathbf{m} = (P, A, D)$ is its PPA metric. The population at generation $t$ is denoted $\mathcal{P}_t = \{\mathcal{I}_1, \ldots, \mathcal{I}_N\}$ with size $N$. The initial population $\mathcal{P}_0$ is seeded using diverse strategies (power-focused, area-focused, timing-focused, balanced, architectural exploration, and simplification) to ensure broad initial coverage. The evolutionary loop iterates for $G$ generations, each comprising offspring generation, evaluation, and survivor selection. Algorithm~\ref{alg:evolution} summarizes the complete procedure.

\begin{algorithm}[t]
\caption{Power-Oriented Evolutionary Optimization}
\label{alg:evolution}
\begin{algorithmic}[1]
\REQUIRE $\mathcal{D}_{\text{orig}}$, $\mathcal{T}$, $N$, $\lambda$, $G$, $R$
\ENSURE Pareto-optimal design set
\STATE $\mathcal{P}_0 \leftarrow \textsc{InitPopulation}(\mathcal{D}_{\text{orig}}, N)$
\FOR{$t = 1$ \TO $G$}
    \STATE $\mathcal{O} \leftarrow \emptyset$
    \FOR{$j = 1$ \TO $\lambda$}
        \STATE Select operator $a_j$ via UCB (Eq.~\ref{eq:ucb})
        \STATE Select parent(s) from $\mathcal{P}_{t-1}$ with $p_i \propto 1/r_i$
        \STATE $\mathcal{D}' \leftarrow \mathrm{LLM}(\text{parent(s)},\, a_j,\, \Delta\mathbf{m})$
        \FOR{$r = 1$ \TO $R$}
            \IF{$\mathcal{T}(\mathcal{D}') = \texttt{PASS}$} \STATE \textbf{break} \ENDIF
            \STATE $\mathcal{D}' \leftarrow \mathrm{LLM}_{\text{repair}}(\mathcal{D}',\, \text{error\_log})$
        \ENDFOR
        \IF{$\mathcal{T}(\mathcal{D}') = \texttt{PASS}$}
            \STATE $\mathbf{m}' \leftarrow \mathrm{Synthesize}(\mathcal{D}')$;~ $\mathcal{O} \leftarrow \mathcal{O} \cup \{(\mathcal{D}', \mathbf{m}')\}$
            \STATE Update UCB reward for $a_j$
        \ENDIF
    \ENDFOR
    \STATE $\mathcal{P}_t \leftarrow \textsc{PowerOrientedSelect}(\mathcal{P}_{t-1} \cup \mathcal{O},\, N)$
\ENDFOR
\RETURN Pareto front of $\mathcal{P}_G$
\end{algorithmic}
\end{algorithm}

\subsubsection{Offspring Generation}\label{subsubsec:offspring}

Each generation produces $\lambda$ offspring by applying evolutionary operators to parents sampled from $\mathcal{P}_{t-1}$ with probability $p_i \propto 1/r_i$, where $r_i$ is the global rank of individual $\mathcal{I}_i$ after power-oriented sorting (Section~\ref{subsubsec:selection}). POET defines six evolutionary operators: five \emph{mutation} operators that each transform a single parent (\textbf{Improve}, \textbf{Refactor}, \textbf{Explore}, \textbf{Simplify}, and \textbf{Fusion}) and one \emph{crossover} operator (\textbf{Crossover}) that combines two parents. All operator prompts include the parent's PPA change relative to $\mathcal{D}_{\text{orig}}$ as a percentage delta $\Delta\mathbf{m}$ to provide quantitative optimization context. Each operator embeds domain-specific circuit optimization techniques, summarized as follows:

\noindent\textbf{Improve} targets the parent's weakest PPA metric from synthesis feedback. \textit{Power}: e.g., clock gating, operand isolation, register update suppression; \textit{Area}: e.g., resource sharing, bit-width reduction, precomputation with LUT replacement; \textit{Delay}: e.g., operator strength reduction, carry optimization, critical path restructuring.

\noindent\textbf{Refactor} keeps the parent's optimization strategy but re-implements it structurally. \textit{Power}: e.g., FSM re-encoding to reduce switching activity; \textit{Area}: e.g., flat logic to parameterized blocks; \textit{Delay \& Area}: e.g., behavioral to structural arithmetic such as Wallace trees.

\noindent\textbf{Explore} generates a maximally different architectural approach. \textit{Power \& Area}: e.g., different encoding schemes or pipeline configurations; \textit{Delay}: e.g., alternative arithmetic structures; \textit{All}: fundamentally different algorithmic alternatives.

\noindent\textbf{Simplify} reduces implementation complexity. \textit{Power}: e.g., eliminating redundant computations; \textit{Area}: e.g., merging redundant logic and reducing bit widths; \textit{Delay}: e.g., simplifying boolean expressions to reduce logic depth.

\noindent\textbf{Fusion} augments a single parent with complementary techniques. \textit{Power + Area}: e.g., clock gating with resource sharing; \textit{Power + Delay}: e.g., operand isolation with strength reduction; \textit{Area + Power}: e.g., bit-width reduction with logic consolidation. Conflicting combinations are explicitly discouraged.

\noindent\textbf{Crossover} combines two parents by inheriting each parent's PPA-superior techniques. \textit{Power}: inherited from the parent with better power; \textit{Area}: from better area; \textit{Delay}: from better delay. Potential conflicts are flagged.

To balance exploration and exploitation across operators, POET employs UCB-based adaptive strategy selection~\cite{qi2024evolution}:
\begin{equation}\label{eq:ucb}
    \mathrm{UCB}(a_i) = \underbrace{\frac{R_i}{n_i}}_{\text{exploitation}} + \; c\;\underbrace{\sqrt{\frac{\ln T}{n_i}}}_{\text{exploration}}
\end{equation}
where $R_i$ and $n_i$ are the cumulative reward and selection count of operator $a_i$, $T$ is the total selection count, and $c$ is an exploration coefficient. The exploitation term favors operators with higher average reward, while the exploration term grows for less frequently selected operators. A reward of $+1$ is granted when the offspring achieves lower power than $\mathcal{D}_{\text{orig}}$.


\begin{table*}[t]
\centering
\caption{PPA comparison on RTL-OPT. Area in $\mu m^2$, CPD in ns, power in $\mu$W. \colorbox{bestgreen}{Green} = best result; \colorbox{failred}{\ding{55}} = functionally incorrect.}
\label{tab:main_results}
\setlength{\tabcolsep}{4pt}
\renewcommand{\arraystretch}{0.6}
\resizebox{\textwidth}{!}{%
\begin{tabular}{l|rrr|rrr|rrr|rrr|rrr}
\toprule
\multirow{2}{*}{Design} & \multicolumn{3}{c|}{Original} & \multicolumn{3}{c|}{I/O} & \multicolumn{3}{c|}{CoT} & \multicolumn{3}{c|}{REvolution} & \multicolumn{3}{c}{POET} \\
\cmidrule(lr){2-4} \cmidrule(lr){5-7} \cmidrule(lr){8-10} \cmidrule(lr){11-13} \cmidrule(lr){14-16}
& Area & CPD & Power & Area & CPD & Power & Area & CPD & Power & Area & CPD & Power & Area & CPD & Power \\
\midrule
add\_sub & 201.89 & 0.58 & 161.0 & 201.89 & 0.58 & 161.0 & 201.89 & 0.58 & 161.0 & \cellcolor{bestgreen}135.93 & 0.64 & 114.0 & 153.75 & \cellcolor{bestgreen}0.55 & \cellcolor{bestgreen}99.20 \\
adder & 458.05 & 1.15 & 393.0 & \cellcolor{failred}\ding{55} & \cellcolor{failred}\ding{55} & \cellcolor{failred}\ding{55} & \cellcolor{failred}\ding{55} & \cellcolor{failred}\ding{55} & \cellcolor{failred}\ding{55} & 409.37 & 1.26 & 363.0 & \cellcolor{bestgreen}272.65 & \cellcolor{bestgreen}1.03 & \cellcolor{bestgreen}195.0 \\
adder\_carry & 52.14 & 0.50 & 32.50 & 48.94 & 0.33 & 29.10 & 48.94 & 0.34 & 29.10 & 47.61 & 0.32 & 28.30 & \cellcolor{bestgreen}47.61 & \cellcolor{bestgreen}0.32 & \cellcolor{bestgreen}28.30 \\
adder\_select & 432.25 & 1.14 & 305.0 & 383.04 & 1.13 & 253.0 & 432.25 & 1.14 & 305.0 & 318.40 & \cellcolor{bestgreen}1.08 & 249.0 & \cellcolor{bestgreen}191.79 & 1.13 & \cellcolor{bestgreen}120.0 \\
addr\_calcu & 124.75 & 0.59 & 196.0 & 124.75 & 0.59 & 196.0 & 124.75 & 0.59 & 196.0 & 124.75 & 0.59 & 196.0 & \cellcolor{bestgreen}118.37 & \cellcolor{bestgreen}0.46 & \cellcolor{bestgreen}138.0 \\
alu\_8bit & 169.44 & 0.44 & 111.0 & 167.85 & 0.47 & 109.0 & 169.44 & 0.44 & 111.0 & \cellcolor{bestgreen}163.06 & 0.48 & 101.0 & 164.65 & \cellcolor{bestgreen}0.39 & \cellcolor{bestgreen}96.40 \\
alu\_64bit & 1483.22 & 3.47 & 1040 & 1392.51 & 2.43 & 983.0 & 1392.51 & \cellcolor{bestgreen}2.43 & 983.0 & \cellcolor{bestgreen}1291.96 & 2.77 & 976.0 & 1321.75 & 2.55 & \cellcolor{bestgreen}944.0 \\
calculation & 902.80 & 2.46 & 1850 & \cellcolor{failred}\ding{55} & \cellcolor{failred}\ding{55} & \cellcolor{failred}\ding{55} & 763.69 & 2.47 & 1680 & 697.72 & 2.48 & 1470 & \cellcolor{bestgreen}657.82 & \cellcolor{bestgreen}2.41 & \cellcolor{bestgreen}1140 \\
comparator & 42.03 & 0.30 & 19.10 & 42.03 & 0.30 & 19.10 & 42.03 & 0.30 & 19.10 & 42.03 & \cellcolor{bestgreen}0.30 & 19.10 & \cellcolor{bestgreen}36.18 & 0.37 & \cellcolor{bestgreen}16.50 \\
comparator\_2bit & 10.11 & 0.16 & 4.73 & 9.31 & 0.08 & 3.73 & 7.98 & 0.09 & 3.47 & 7.98 & 0.08 & 3.47 & \cellcolor{bestgreen}7.98 & \cellcolor{bestgreen}0.08 & \cellcolor{bestgreen}3.47 \\
comparator\_4bit & 21.55 & 0.16 & 9.99 & 17.02 & \cellcolor{bestgreen}0.12 & 8.19 & 17.82 & 0.14 & 8.32 & 12.77 & 0.16 & 4.87 & \cellcolor{bestgreen}12.77 & 0.16 & \cellcolor{bestgreen}4.87 \\
comparator\_8bit & 47.88 & 0.24 & 20.90 & 30.32 & 0.22 & 14.60 & 34.85 & \cellcolor{bestgreen}0.20 & 16.10 & 30.32 & 0.22 & 14.60 & \cellcolor{bestgreen}30.32 & 0.22 & \cellcolor{bestgreen}14.60 \\
comparator\_16bit & 104.80 & 0.29 & 44.50 & 65.17 & 0.29 & 32.80 & 67.83 & 0.28 & 31.20 & 36.71 & 0.22 & 14.30 & \cellcolor{bestgreen}36.71 & \cellcolor{bestgreen}0.22 & \cellcolor{bestgreen}14.30 \\
decoder\_6bit & 99.22 & 0.21 & 19.50 & 99.22 & 0.21 & 19.50 & 99.22 & 0.21 & 19.50 & 78.20 & 0.11 & 19.50 & \cellcolor{bestgreen}78.20 & \cellcolor{bestgreen}0.11 & \cellcolor{bestgreen}19.50 \\
decoder\_8bit & 410.70 & 0.48 & 57.30 & 423.21 & 0.63 & 58.10 & 422.41 & 1.14 & 102.0 & 250.04 & 0.18 & 46.60 & \cellcolor{bestgreen}250.04 & \cellcolor{bestgreen}0.18 & \cellcolor{bestgreen}46.60 \\
divider\_4bit & 40.96 & 0.34 & 24.60 & 40.96 & 0.34 & 24.60 & 40.96 & 0.34 & 24.60 & 35.91 & \cellcolor{bestgreen}0.31 & 21.20 & \cellcolor{bestgreen}24.21 & 0.35 & \cellcolor{bestgreen}11.10 \\
divider\_8bit & 193.65 & \cellcolor{bestgreen}1.42 & 508.0 & \cellcolor{failred}\ding{55} & \cellcolor{failred}\ding{55} & \cellcolor{failred}\ding{55} & \cellcolor{failred}\ding{55} & \cellcolor{failred}\ding{55} & \cellcolor{failred}\ding{55} & 178.75 & 1.49 & 407.0 & \cellcolor{bestgreen}178.75 & 1.49 & \cellcolor{bestgreen}407.0 \\
divider\_16bit & 1142.47 & 6.35 & 102000 & 1142.47 & 6.35 & 102000 & 752.51 & 5.44 & 26700 & 754.11 & 5.61 & 24700 & \cellcolor{bestgreen}748.52 & \cellcolor{bestgreen}5.42 & \cellcolor{bestgreen}21400 \\
divider\_32bit & 3196.79 & 18.28 & 275000 & \cellcolor{failred}\ding{55} & \cellcolor{failred}\ding{55} & \cellcolor{failred}\ding{55} & 3196.79 & \cellcolor{bestgreen}18.28 & 275000 & 3028.14 & 21.25 & 241000 & \cellcolor{bestgreen}3028.14 & 21.25 & \cellcolor{bestgreen}241000 \\
fsm & 111.45 & 0.24 & 123.0 & \cellcolor{failred}\ding{55} & \cellcolor{failred}\ding{55} & \cellcolor{failred}\ding{55} & 160.66 & 0.48 & 50.70 & \cellcolor{bestgreen}77.67 & \cellcolor{bestgreen}0.22 & 71.60 & 113.32 & 0.33 & \cellcolor{bestgreen}26.80 \\
fsm\_encode & \cellcolor{bestgreen}332.23 & \cellcolor{bestgreen}0.62 & 1150 & \cellcolor{failred}\ding{55} & \cellcolor{failred}\ding{55} & \cellcolor{failred}\ding{55} & \cellcolor{failred}\ding{55} & \cellcolor{failred}\ding{55} & \cellcolor{failred}\ding{55} & \cellcolor{failred}\ding{55} & \cellcolor{failred}\ding{55} & \cellcolor{failred}\ding{55} & 381.18 & 0.65 & \cellcolor{bestgreen}226.0 \\
gray & \cellcolor{bestgreen}99.22 & 0.32 & 384.0 & 119.43 & \cellcolor{bestgreen}0.23 & 167.0 & 155.08 & 0.38 & 242.0 & 161.20 & 0.37 & 117.0 & 261.21 & 0.86 & \cellcolor{bestgreen}102.0 \\
mac & 1020.11 & 0.99 & 2800 & \cellcolor{failred}\ding{55} & \cellcolor{failred}\ding{55} & \cellcolor{failred}\ding{55} & \cellcolor{failred}\ding{55} & \cellcolor{failred}\ding{55} & \cellcolor{failred}\ding{55} & 975.69 & 0.92 & 1940 & \cellcolor{bestgreen}975.69 & \cellcolor{bestgreen}0.92 & \cellcolor{bestgreen}1940 \\
mul & 496.89 & 0.87 & 1280 & \cellcolor{failred}\ding{55} & \cellcolor{failred}\ding{55} & \cellcolor{failred}\ding{55} & \cellcolor{failred}\ding{55} & \cellcolor{failred}\ding{55} & \cellcolor{failred}\ding{55} & \cellcolor{bestgreen}486.25 & 0.80 & 1080 & 491.04 & \cellcolor{bestgreen}0.78 & \cellcolor{bestgreen}1060 \\
mul\_const & 69.69 & 0.26 & 47.00 & 52.40 & 0.21 & 30.30 & 52.40 & 0.21 & 30.30 & 52.40 & 0.21 & 30.30 & \cellcolor{bestgreen}52.40 & \cellcolor{bestgreen}0.21 & \cellcolor{bestgreen}30.30 \\
mul\_subexpression & 345.27 & 1.23 & 1400 & 506.46 & 1.53 & 819.0 & 364.15 & 0.83 & 596.0 & 339.68 & 0.78 & 489.0 & \cellcolor{bestgreen}335.69 & \cellcolor{bestgreen}0.72 & \cellcolor{bestgreen}443.0 \\
multi\_if & 10.91 & 0.15 & 4.21 & 7.98 & 0.10 & 3.18 & 10.91 & 0.15 & 4.21 & \cellcolor{failred}\ding{55} & \cellcolor{failred}\ding{55} & \cellcolor{failred}\ding{55} & \cellcolor{bestgreen}7.98 & \cellcolor{bestgreen}0.10 & \cellcolor{bestgreen}3.18 \\
mux\_4to1\_16bit & 83.79 & 0.19 & 29.60 & 83.79 & 0.19 & 29.60 & \cellcolor{bestgreen}83.79 & 0.19 & 29.60 & 89.38 & 0.11 & 29.30 & 89.38 & \cellcolor{bestgreen}0.11 & \cellcolor{bestgreen}29.30 \\
mux\_4to1\_64bit & 326.38 & 0.60 & 138.0 & 326.38 & 0.60 & 138.0 & \cellcolor{bestgreen}326.38 & 0.59 & 138.0 & 357.50 & 0.11 & 117.0 & 357.50 & \cellcolor{bestgreen}0.11 & \cellcolor{bestgreen}117.0 \\
mux\_dead & 21.28 & 0.04 & 7.06 & 21.28 & 0.04 & 7.06 & 21.28 & 0.04 & 7.06 & 21.28 & 0.04 & 7.06 & \cellcolor{bestgreen}21.28 & \cellcolor{bestgreen}0.04 & \cellcolor{bestgreen}7.06 \\
mux\_encode & 63.57 & 0.21 & 23.80 & 63.57 & 0.21 & 23.80 & 63.57 & 0.21 & 23.80 & 63.57 & 0.20 & 23.80 & \cellcolor{bestgreen}63.57 & \cellcolor{bestgreen}0.20 & \cellcolor{bestgreen}23.50 \\
mux\_large & 124.49 & 0.34 & 53.00 & 90.17 & 0.30 & 29.90 & 90.17 & 0.30 & 29.90 & 92.57 & 0.30 & 29.30 & \cellcolor{bestgreen}88.58 & \cellcolor{bestgreen}0.30 & \cellcolor{bestgreen}28.60 \\
register & \cellcolor{bestgreen}9712.19 & \cellcolor{bestgreen}0.88 & 14300 & \cellcolor{failred}\ding{55} & \cellcolor{failred}\ding{55} & \cellcolor{failred}\ding{55} & \cellcolor{failred}\ding{55} & \cellcolor{failred}\ding{55} & \cellcolor{failred}\ding{55} & \cellcolor{failred}\ding{55} & \cellcolor{failred}\ding{55} & \cellcolor{failred}\ding{55} & 10143.38 & 1.07 & \cellcolor{bestgreen}8290 \\
saturating\_add & 78.47 & 0.26 & 83.60 & \cellcolor{failred}\ding{55} & \cellcolor{failred}\ding{55} & \cellcolor{failred}\ding{55} & 78.47 & 0.26 & 83.60 & 42.29 & 0.26 & 25.00 & \cellcolor{bestgreen}42.29 & \cellcolor{bestgreen}0.26 & \cellcolor{bestgreen}25.00 \\
selector & 42.29 & 0.12 & 37.80 & 37.24 & 0.11 & 36.60 & 37.24 & 0.11 & 36.60 & 10.11 & 0.11 & 4.30 & \cellcolor{bestgreen}10.11 & \cellcolor{bestgreen}0.11 & \cellcolor{bestgreen}4.30 \\
sub\_4bit & 22.34 & 0.17 & 11.50 & 22.34 & 0.17 & 11.40 & \cellcolor{failred}\ding{55} & \cellcolor{failred}\ding{55} & \cellcolor{failred}\ding{55} & 20.75 & \cellcolor{bestgreen}0.17 & 11.30 & \cellcolor{bestgreen}20.48 & 0.23 & \cellcolor{bestgreen}10.70 \\
sub\_8bit & 48.15 & 0.34 & 28.80 & 48.15 & 0.34 & 29.40 & 48.15 & 0.34 & 28.80 & 48.15 & 0.34 & 28.80 & \cellcolor{bestgreen}46.28 & \cellcolor{bestgreen}0.33 & \cellcolor{bestgreen}27.40 \\
sub\_16bit & 102.14 & 0.62 & 63.60 & 98.95 & 0.57 & 63.50 & 98.95 & 0.57 & 63.40 & 97.89 & 0.64 & 61.80 & \cellcolor{bestgreen}96.56 & \cellcolor{bestgreen}0.57 & \cellcolor{bestgreen}58.10 \\
sub\_32bit & 208.54 & 1.31 & 137.0 & \cellcolor{failred}\ding{55} & \cellcolor{failred}\ding{55} & \cellcolor{failred}\ding{55} & 202.16 & 1.22 & 130.0 & \cellcolor{failred}\ding{55} & \cellcolor{failred}\ding{55} & \cellcolor{failred}\ding{55} & \cellcolor{bestgreen}196.57 & \cellcolor{bestgreen}1.10 & \cellcolor{bestgreen}121.0 \\
ticket\_machine & 59.32 & 0.39 & 75.70 & 57.72 & 0.35 & 62.90 & 57.72 & 0.35 & 62.90 & \cellcolor{bestgreen}33.25 & 0.26 & 37.90 & 40.43 & \cellcolor{bestgreen}0.20 & \cellcolor{bestgreen}29.10 \\
\bottomrule
\end{tabular}%
}
\end{table*}

\subsubsection{Evaluation and Repair}\label{subsubsec:eval}

Each offspring $\mathcal{D}'$ is verified against the checking testbench $\mathcal{T}$. If $\mathcal{T}(\mathcal{D}') \neq \texttt{PASS}$, the error log is fed back to the LLM for repair, repeating for up to $R$ attempts; offspring that still fail are discarded. This maintains an \emph{all-correct population invariant}: every individual in $\mathcal{P}_t$ satisfies $\mathcal{T}(\mathcal{I}) = \texttt{PASS}$. Verified offspring are then evaluated for PPA through logic synthesis.

\subsubsection{Power-Oriented Survivor Selection}\label{subsubsec:selection}

After evaluation, the combined pool $\mathcal{P}_{t-1} \cup \mathcal{O}$ is reduced to $N$ survivors through three steps built upon NSGA-II~\cite{deb2002nsga2}:

\noindent\textbf{(1) Non-dominated sorting.} The pool is partitioned into Pareto levels $\mathcal{F}_1, \mathcal{F}_2, \ldots, \mathcal{F}_L$ via the dominance relation $\succ$ (Eq.~\ref{eq:dominance}), where $\mathcal{F}_1$ is the Pareto front and subsequent levels represent lower quality.

\noindent\textbf{(2) Intra-level ranking.} Within each level $\mathcal{F}_k$, individuals are sorted by power in ascending order via $\succ_P$ (Eq.~\ref{eq:power_rank}), ensuring lower-power designs rank higher among Pareto-equivalent candidates.

\noindent\textbf{(3) Proportional slot allocation.} Unlike standard NSGA-II sequential fill, which may completely eliminate lower Pareto levels, POET assigns each level a quota proportional to its priority:
\begin{equation}\label{eq:quota}
    s_k = \max\!\left(1,\; \lfloor N \cdot w_k \rfloor\right), \quad w_k = \frac{L - k + 1}{\sum_{j=1}^{L}(L - j + 1)}
\end{equation}
Higher-priority levels (lower $k$) receive larger quotas, while every level retains at least one representative to preserve potentially valuable optimization strategies. Individuals are selected first by Pareto level priority, then by power priority within each level.

\section{Experiments}\label{sec:experiments}

We conduct comprehensive experiments to answer three research questions: \textbf{RQ1}: How does POET compare with existing LLM-based RTL optimization methods in functional correctness and PPA quality? \textbf{RQ2}: What is the contribution of each core component? \textbf{RQ3}: How does the evolutionary process optimize PPA over generations?

\subsection{Experimental Setup}\label{subsec:setup}

\noindent\textbf{Benchmark and Baselines.} We evaluate on the RTL-OPT benchmark~\cite{lu2026benchmark}, which contains 40 expert-crafted RTL designs spanning arithmetic circuits, control logic, and datapath modules. Since RTL-OPT provides two versions per design, we select the one with higher power as the optimization target to align with our power-centric objective. We compare POET against three baselines: (1)~\textbf{I/O prompting}, which directly instructs the LLM to optimize the RTL code; (2)~\textbf{Chain-of-Thought (CoT)}~\cite{wei2022chain}, which guides the LLM through step-by-step optimization reasoning; and (3)~\textbf{REvolution}~\cite{min2025revolution}, a state-of-the-art evolutionary framework using weighted-sum fitness. All methods use GPT-4o-mini as the backbone LLM with the same total number of LLM generation calls for fair comparison.

\begin{figure*}[t]
\centering
\includegraphics[width=0.95\textwidth]{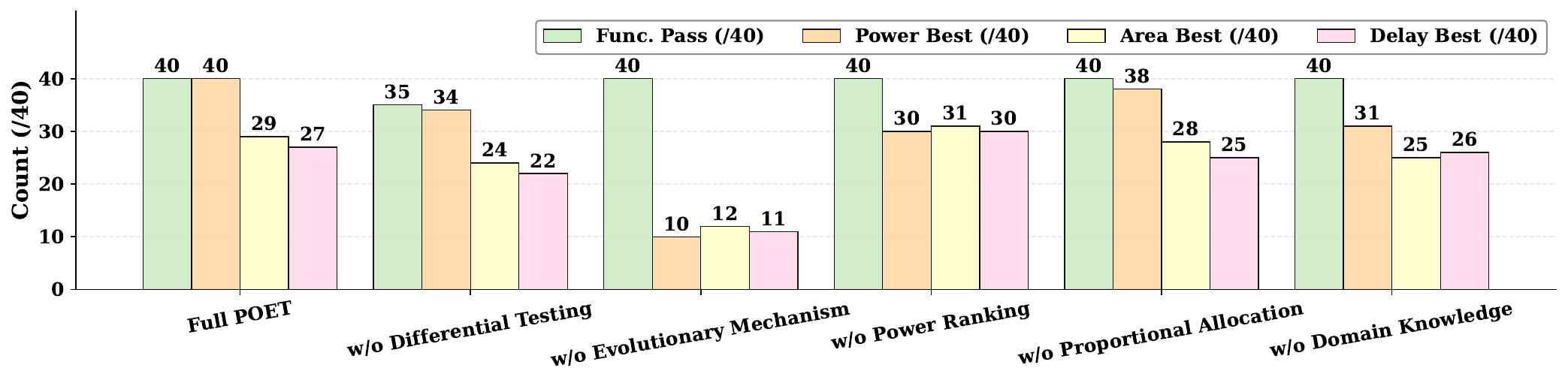}
\caption{Ablation study on RTL-OPT. Each bar indicates the number of designs achieving the best result among all baselines.}
\label{fig:ablation}
\end{figure*}

\begin{figure}[t]
\centering
\includegraphics[width=0.95\columnwidth]{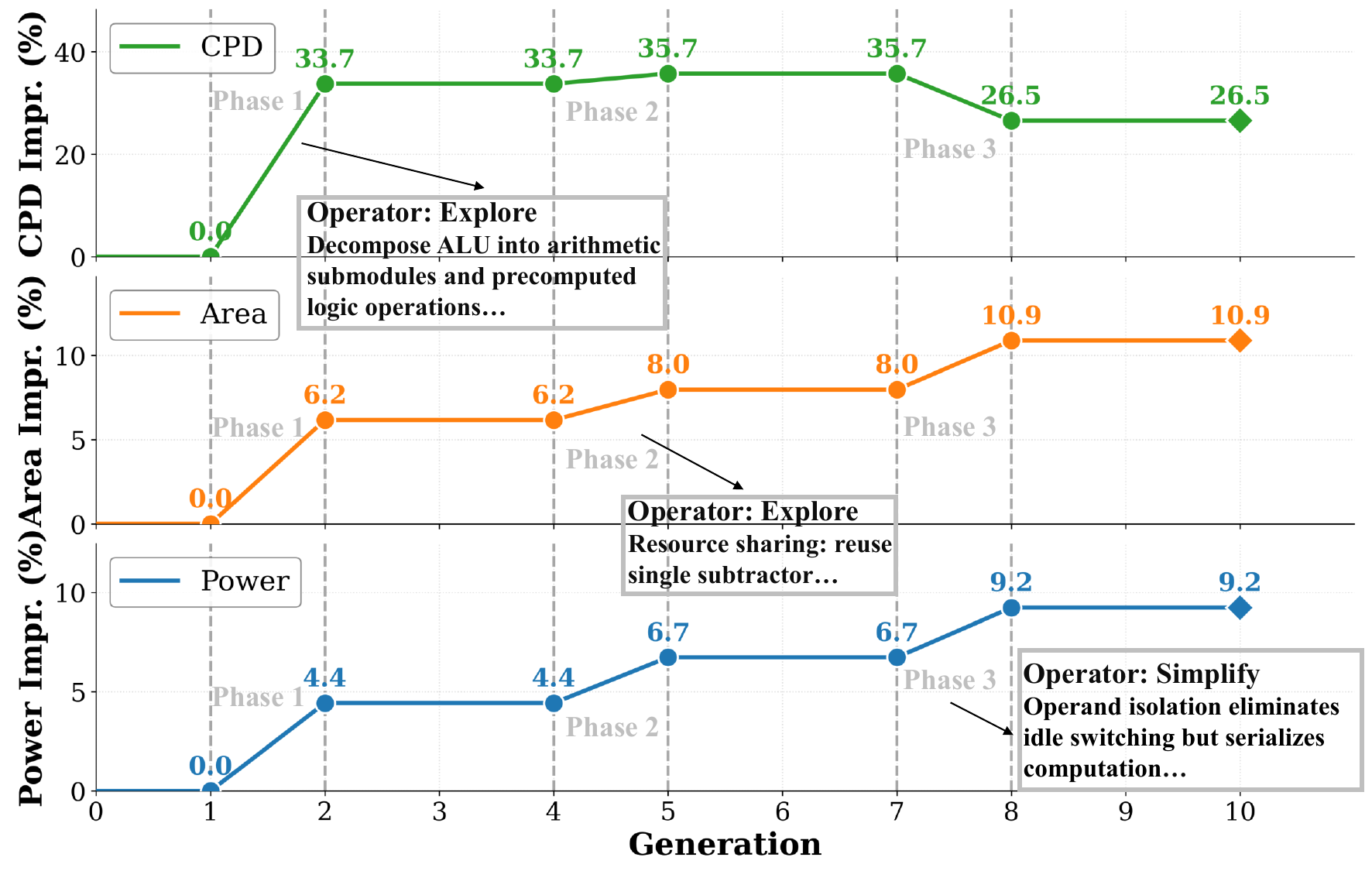}
\caption{Evolutionary trajectory of POET on \texttt{alu\_64bit}.}
\label{fig:case_study}
\end{figure}

\noindent\textbf{Evaluation.} PPA metrics (area in $\mu m^2$, critical path delay in ns, power in $\mu$W) are obtained using Yosys for logic synthesis and OpenSTA for timing and power analysis with the NanGate 45nm standard cell library. Functional correctness is verified by both simulation via Icarus Verilog against POET's generated testbenches (manually inspected for reliability) and Yosys equivalence checking. Since RTL-OPT does not provide pre-built testbenches, REvolution uses LLM-generated testbenches for its internal verification. 

\noindent\textbf{Hyperparameters.} POET uses population size $N{=}10$, offspring count $\lambda{=}10$, maximum generations $G{=}10$, repair attempts $R{=}3$.

\subsection{Main Results (RQ1)}\label{subsec:main_results}

Table~\ref{tab:main_results} presents the per-design PPA comparison. POET achieves the best power on all 40 designs (100\%), the best area on 29 (72.5\%), and the best delay on 27 (67.5\%), while being the only method to produce functionally correct results for all 40 benchmarks.

\noindent\textbf{Functional Correctness.} The differential testing pipeline enables POET to achieve 100\% functional correctness (40/40), compared to 72.5\% for I/O prompting (29/40), 82.5\% for CoT (33/40), and 90.0\% for REvolution (36/40). I/O, CoT, and REvolution fail on 11, 7, and 4 designs respectively, with failures concentrated on complex arithmetic and sequential circuits. For \texttt{fsm\_encode} and \texttt{register}, where all three baselines fail, POET is the only method that succeeds by leveraging its differential testing pipeline to generate reliable testbenches from the original design, enabling comprehensive verification of the restructured state encoding and register access patterns.

\noindent\textbf{Power-Centric PPA Optimization.} POET achieves the best power on every design, demonstrating the effectiveness of the power-oriented evolutionary mechanism. The results highlight two key optimization capabilities of POET. First, when the design is not yet Pareto-optimal, POET improves all PPA metrics simultaneously. For example, \texttt{ticket\_machine} achieves 31.8\% area, 48.7\% CPD, and 61.6\% power reduction over the original, and \texttt{comparator\_16bit} achieves 65.0\% area, 24.1\% CPD, and 67.9\% power reduction over the original. Second, when the design is near the Pareto front, POET further reduces power through controlled trade-offs. For \texttt{comparator}, POET reduces power by 13.6\% and area by 13.9\% over the original while CPD increases slightly from 0.30\,ns to 0.37\,ns; for \texttt{gray}, POET achieves the lowest power (102.0$\mu$W, 12.8\% below the best baseline) at the cost of increased area.


\subsection{Ablation Study (RQ2)}\label{subsec:ablation}

We evaluate five ablation configurations, each removing one component from the full POET framework: (1)~\textbf{w/o Differential Testing (DT)}, replacing the differential testing pipeline with LLM-generated testbenches; (2)~\textbf{w/o Evolutionary Mechanism (EM)}, replacing evolutionary survivor selection with random selection; (3)~\textbf{w/o Power Ranking (PR)}, using random intra-level selection instead of power-ascending ranking; (4)~\textbf{w/o Proportional Allocation (PA)}, using standard NSGA-II sequential fill; and (5)~\textbf{w/o Domain Knowledge (DK)}, removing circuit-specific techniques from operator prompts. Figure~\ref{fig:ablation} reports the Best count for each setting, where each bar indicates the number of designs (out of 40) achieving the best result among all baselines (Original, I/O, CoT, REvolution).

\textbf{w/o DT} is the only setting that degrades functional correctness, with Func. Pass dropping from 40 to 35, which cascades into reduced Power Best (34), Area (24), and Delay (22) as unreliable LLM-generated testbenches compromise verification throughout evolution. \textbf{w/o EM} causes the largest degradation across all metrics (Power: 40$\to$10, Area: 29$\to$12, Delay: 27$\to$11), confirming that evolutionary selection is the fundamental driver of optimization. \textbf{w/o PR} drops Power Best from 40 to 30 while Area and Delay Best shift to 31 and 30, revealing that without power-oriented ranking the search redistributes capacity to other metrics. \textbf{w/o PA} shows a smaller drop (Power: 40$\to$38), indicating proportional allocation provides a consistent but moderate benefit through diversity preservation. \textbf{w/o DK} reduces Power Best to 31, Area to 25, and Delay to 26, demonstrating that hardware-specific techniques in operator prompts provide meaningful guidance for RTL transformations.

\subsection{Case Study (RQ3)}\label{subsec:case_study}

Figure~\ref{fig:case_study} traces POET's optimization trajectory on \texttt{alu\_64bit} across 10 generations, where improvements are measured relative to the original design.
In Phase~1 (generation~2), \textbf{Explore} decomposes the monolithic ALU into specialized submodules, achieving 33.7\% CPD, 6.2\% area, and 4.4\% power reduction. In Phase~2 (generation~5), \textbf{Explore} applies resource sharing, further improving to 8.0\% area and 6.7\% power reduction while CPD reaches 35.7\%. In Phase~3 (generation~8), \textbf{Simplify} applies operand isolation to eliminate idle switching activity, pushing power to 9.2\% and area to 10.9\%, while CPD settles at 26.5\% due to serialization overhead.
This trajectory exemplifies two capabilities of POET: collectively improving all PPA metrics through complementary operators (Phases~1 and~2), and further reducing power through controlled trade-offs near the Pareto front at modest CPD cost (Phase~3).

\section{Conclusion}\label{sec:conclusion}

We presented POET, a framework for power-centric RTL PPA optimization that addresses two fundamental challenges: functional correctness and multi-objective optimization directionality. The differential-testing-based testbench generation pipeline leverages the original design as a functional oracle, producing reliable verification through deterministic simulation rather than LLM reasoning. The power-oriented evolutionary mechanism combines non-dominated sorting, power-first intra-level ranking, and proportional survivor selection to systematically steer the search toward low-power solutions while preserving Pareto optimality, without manual weight tuning. Experiments on RTL-OPT demonstrate that POET achieves 100\% functional correctness and the best power on all 40 designs, with competitive area and delay improvements. Ablation studies and a case study further validate each component's contribution.



\bibliographystyle{ACM-Reference-Format}
\bibliography{references}


\end{document}